\documentclass[12pt,a4paper]{article}
\pdfoutput=1
\usepackage{jheppub}
\usepackage[T1]{fontenc} 
\usepackage{epsfig} 
\usepackage{amscd}
\usepackage{verbatim}
\usepackage{latexsym}
\usepackage{amsfonts,amsthm,amsmath,mathtools}
\usepackage{amssymb,stmaryrd,textcomp,enumerate,bbm}
\usepackage{color}
\usepackage{xcolor}
\usepackage{booktabs}
\usepackage{float}
\usepackage{graphicx}
\usepackage{slashed}

\notoc 

\title{\centering{SUSY Breaking  in Monopole Quivers 
}}

\author[a]{Antonio Amariti}

\affiliation[a]{INFN Sezione di Milano, Via Celoria 16, Milano, Italy}

\emailAdd
{antonio.amariti@mi.infn.it} 

\newcommand{\nocontentsline}[3]{}
\newcommand{\tocless}[2]{\bgroup\let\addcontentsline=\nocontentsline#1{#2}\egroup}

\abstract{
We claim that 3d monopole quivers, theories with product gauge groups interacting through Affleck-Harvey-Witten 
superpotentials, are a natural setup for the study of
spontaneous breaking 3d $\mathcal{N}=2$ supersymmetry.
We give evidence of this statement by studying various examples of increasing complexity, 
considering quivers that are mirror dual to Wess-Zumino models.
These Wess-Zumino models, in opportune regimes of parameters, break supersymmetry in
perturbatively controllable (meta)stable vacua.
}

\begin{document}

\maketitle

\newpage
\tableofcontents
\newpage
\section{Introduction}
%
%
%
%
%
%

In the recent past the three dimensional version of 2d
bosonization attracted a large interest (see for example 
 \cite{Giombi:2011kc,Aharony:2011jz,Aharony:2012nh,Aharony:2015mjs,Seiberg:2016gmd}). 
This is just the tip of
a larger web of 3d dualities, sharing many similarities with the
supersymmetric case. 
The possibility of deriving non-supersymmetric dualities from supersymmetry
 has been  explored in \cite{Jain:2013gza,Gur-Ari:2015pca,Kachru:2016rui,Kachru:2016aon,Aharony:2018pjn},
 by explicit but controllable SUSY breaking.

An intriguing, but so far unexplored possibility, consists of deriving 
3d non-supersymmetric dualities from spontaneous or dynamical supersymmetry breaking.
Examples of $\mathcal{N}=2 \rightarrow \mathcal{N}=0$ breaking in
3d gauge theories  have been discussed in 
\cite{Witten:1999ds,Bergman:1999na,Ohta:1999iv,deBoer:1997kr},
mostly using arguments from the brane dynamics,  even if a complete field theoretical analysis 
is missing.

Another interesting case was discussed in \cite{Giveon:2009bv}, generalizing the ISS mechanism
\cite{Intriligator:2006dd}  to 3d $\mathcal{N}=2$ $U(N_c)_k$
SQCD,  where $k$ is the Chern-Simons (CS) level.
In the IR the theory reduces to a Wess-Zumino (WZ)   model, that, in a proper regime of 
couplings, gives perturbatively accessible, long lived and non-supersymmetric metastable vacua.
Furthermore 3d supersymmetry breaking for  WZ models has been explored
in the literature in \cite{Amariti:2009kb}. It has been observed that supersymmetry 
breaking can occur for large classes of models, with both classically marginal and relevant
deformations, and that these vacua are often (long lived) metastable ones.

In this paper we observe that many of the models discussed in \cite{Giveon:2009bv,Amariti:2009kb} can be derived
from new types of quiver gauge theories, 
defined in \cite{Amariti:2017gsm} as monopole quivers.
Differently from common quivers, here the gauge nodes are connected through
Affleck-Harvey-Witten (AHW) superpotential interactions \cite{Affleck:1982as}, involving monopole operators.

Here we will consider monopole quivers with 
$U(1)_k$ gauge factors with charged matter fields
and deform some of the nodes by linear or quadratic monopole superpotentials,
of the type discussed in \cite{Benini:2017dud,Amariti:2018gdc}.
By studying a large variety of examples we show that the $\mathcal{N}=2 \rightarrow \mathcal{N}=0$ breaking in monopole quivers
is quite generic.
The analysis is possible after performing 
mirror symmetry at the various nodes, 
dualizing the monopole quivers to WZ models.
Many of these WZ models reduce in the IR to the ones discussed in \cite{Giveon:2009bv,Amariti:2009kb},
that have been shown to break supersymmetry in either stable or long lived metastable 
non-supersymmetric vacua.

The paper is organized as follows.
In section \ref{sec:review} we collect some review material, useful for
our analysis.
In section \ref{sec:exDSB} we discuss various examples  of monopole quivers leading to (meta)stable supersymmetry breaking.
In section \ref{ref:conclusions} we discuss open questions and propose possible generalizations of our analysis.
In appendix \ref{appendix} we discuss the one loop effective potential
in presence of spontaneous supersymmetry breaking and  the bounce action for a 3d triangular barrier, necessary for the estimation of the lifetime of the supersymmetry breaking vacuum.
In appendix \ref{appB} and in appendix  \ref{appC} we give e detailed analysis of supersymmetry
breaking for two WZ models obtained in section \ref{sec:exDSB} that, to our knowledge, have never been analyzed in the literature.

\section{Review}
\label{sec:review}

We will not review here generic aspects of 3d $\mathcal{N}=2$ 
gauge theories. We refer the reader to \cite{Aharony:1997bx} for definitions and general discussions.
Here we restrict our attention to
supersymmetry breaking in 3d WZ models,
to 3d dualities and to the definition of monopole quivers.

\subsection{Supersymmetry breaking in 3d WZ}
\label{revSB}
In this section we overview some of the results on
supersymmetry breaking in 3d WZ models
obtained in \cite{Giveon:2009bv,Amariti:2009kb}.
We refer the reader to the original references for
more complete discussions.
We collect in the Appendix \ref{appendix} the various tools for the analysis of the 1-loop effective potential and for the 
estimation of the lifetime of the metastable state through the evaluation of the bounce action for a triangular barrier.

In our analysis we consider WZ models of the form
\begin{equation}
W_{WZ} =
f X + \sum_{i=0}^{2} \sum_{a,b=1}^n X^i M^{(i)}_{a b } \phi_a \phi_b 
\end{equation}
where there are $n+1$ chiral superfields $X$ and 
$\phi_i$ and the 
couplings are encoded in the three matrices $M^{(i)}$.
A classification scheme for supersymmetry breaking 
can be constructed by separating the cases with 
either $M^{(1)}=0$ (marginal couplings) or $M^{(2)}=0$ (relevant couplings).

The first cases has been fully classified in \cite{Amariti:2009kb} and we will report here the main results.
The second case can be further divided in sub-classes, depending on the possible $R$-charge assignations.
We will discuss this classifications distinguishing the various possibilities.
There is a third case, where neither $M^{(1)} = 0$ nor 
$M^{(2)} = 0$. We will comment about this case as well.

\tocless\subsubsection{Marginal couplings}

This family has been analyzed in \cite{Amariti:2009kb}.
The main results of the analysis are
\begin{itemize}
\item The perturbative analysis is reliable if the marginal
coupling constants encoded in $M^{(2)}$ are small numbers.
\item The origin is a local maximum. This implies that the 
non-supersymmetric vacuum spontaneously breaks $U(1)_R$.
\item The scalar potential has a classical runaway behavior.
\end{itemize}
~
\\
{\bf A case study}. The simplest example of this family is given by the superpotential 
\begin{equation}
\label{rel}
W = f X + h X^2 \phi_1^2 + \mu \phi_1 \phi_2
\end{equation}
where the fields are considered as free and the mass dimensions of the couplings are 
$[h]=0$,  $[\mu] =1$ and $[f]= \frac{3}{2}$.
Next we summarize the detailed analysis of \cite{Amariti:2009kb} for 
this case, because 
it will play a crucial role here.
First it is necessary to compute the F-terms
\begin{equation}
F_X =  f + 2 h X \phi_1^2 = 0,
\quad
F_{\phi_1} = 2 h X^2 \phi_1+ \mu \phi_2 = 0
, \quad
F_{\phi_2} = \mu \phi_1 = 0
\end{equation}
and observe that they cannot be solved
simultaneously, signaling that
supersymmetry is broken at tree level.
In order to have a trustable supersymmetry breaking model 
one needs to study the quantum corrections around the
tree level supersymmetry breaking minimum.
The tree level scalar potential
\footnote{Assuming a canonical K\"ahler potential.} 
is
\begin{equation}
V_{tree} = |F_X|^2 + |F_{\phi_1}|^2 +  |F_{\phi_2}|^2
\end{equation}
and there is a classical flat direction associated to the fields  $X$. It identifies the supersymmetry breaking locus
together with $\phi_1= \phi_2 =0$.
This locus is stable at tree level if the squared masses of the 
scalars in the superfields $\phi_i$ are positive.
The four masses of the real bosonic components of the super-fields $\phi_i$ are
\begin{equation}
m_B^2 = \mu^2 + 2 h X (h X^3 + \eta_1 f + 
\eta_2 \sqrt
{ f^2+2 \eta f h X^3 + h^2 X^6 + X^2 \mu^2}
)
\end{equation}
with both $\eta_i = \pm 1$
and they are positive for 
\begin{equation}
\label{tfree}
4 f X < \mu^2
\end{equation}
The fermionic masses are obtained by setting $f=0$ in $m_B$.
The scalar component of the superfield $X$ is vanishing, corresponding to 
a pseudomodulus, i.e. an accidental flat direction of the tree level scalar potential
that can be lifted at quantum level.
The two real fermionic combinations of the superfield $X$ correspond to the  two goldstinos of the
$\mathcal{N} = 2 \rightarrow \mathcal{N} = 0$ supersymmetry breaking.
The knowledge of the massive  tree level spectrum of 
the supersymmetry breaking locus can be used to study the quantum correction, 
through the Coleman-Weinberg (CW)  potential
\cite{Coleman:1973jx}.
 This is the effective potential that can lift
the flat direction $X$.
In 3d the one loop contribution to the CW effective potential is given by (\ref{CW}).
By performing the calculation in our case we observe that the origin is tachyonic, $m_{X=0}^2 = -(f h)^2/m$, while there is 
a minimum at $X \simeq 2^{1/4} \sqrt{\mu/h}$, if $f^2 \ll \mu^3$.   
Furthermore condition (\ref{tfree}) imposes 
\begin{equation}
 h \ll \frac{1}{16 \sqrt 2} \frac{\mu^3}{f^2} 
\end{equation}
This signals the fact that 
the perturbative expansion is reliable
because we are free to chose the dimensionless coupling $h$ to be small enough to suppress higher loops.

The stability of this non-supersymmetric vacuum 
has to be still checked
by studying the behavior of the tree level scalar potential
at large vev.
By parameterizing the fields as
\begin{equation}
X =\frac{f}{2 h m} e^{2 \alpha},
\quad 
\phi_1 = \sqrt m e^{-\alpha},
\quad 
\phi_3 = \frac{f}{2 h \sqrt m^5} e^{3\alpha}
\end{equation}
we have that $V_{tree} \rightarrow 0$ 
in the limit $\alpha \rightarrow 0$.
The presence of this runway signals that the non-supersymmetric state is only metastable.
The lifetime of this metastable state is studied by estimating the bounce action $S_B$ 
for a triangular barrier. The decay rate of the state is given by $e^{-S_B}$.
We have (see appendix \ref{appendix} for details)
\begin{equation}
\label{eq:bouncemarg}
S_B  \propto \sqrt{\frac{(\Delta \Phi)^6}{\Delta V}}
\propto \left(\frac{\mu^3}{f^2}\right)^2
\sqrt{\frac{1}{h}} \gg
1
\end{equation}
where, because of the presence of the runaway,
$\Delta \Phi$ is estimated starting at $X_{min}$
and ending at a value $X=X^* \neq X_{min}$ such that  
$V(X^*) = V_{min}$. The denominator in 
(\ref{eq:bouncemarg}) is
$\Delta V = V_{max} - V_{min}$.

In section \ref{sec:exDSB} we will find many quiver gauge theories that
in the IR reduce to the WZ model with the superpotential
given  in formula (\ref{rel}).
In all these cases a long lived supersymmetry breaking 
metastable minimum exists if the various parameters satisfy the constraints discussed here.

\tocless\subsubsection{Relevant couplings}

Models with relevant couplings give raise to a quite different 
analysis. Indeed, as observed in \cite{Giveon:2009bv}, 
there is a problem with the reliability of the 
perturbative expansion in these cases.
This issue was solved in \cite{Giveon:2009bv} by adding an explicit R-symmetry breaking deformation
\footnote{
Observe that, depending on the structure of the 
superpotential, there can be models with relevant coupling leading to spontaneous $R$-symmetry breaking.
In such cases the requirement of an explicit R-symmetry breaking deformation is not always necessary.}.
The case discussed in
\cite{Giveon:2009bv} corresponds to a WZ model with  superpotential 
\begin{equation}
\label{marg}
W = \frac{1}{2} h \mu \epsilon  X^2 
- h \mu^2 X 
+h \mu \epsilon \phi_3 \phi_4 
 +h X \phi_1 \phi_2 
 +h \mu (\phi_1 \phi_3+\phi_2 \phi_4) 
\end{equation}
where the fields are considered as free and the mass dimensions of the couplings are $[h] =[\mu] = \frac{1}{2}$ and $[\epsilon]=0$.
This theory has a non-supersymmetric vacuum at
$\phi_i=0$ and $X \simeq \epsilon \mu b$, with
$b \equiv \hat b \frac{\mu}{h} \equiv \frac{4 \pi \mu }{(3 - 2 \sqrt 2)h}$, dimensionless.
This vacuum is close to the origin if $\epsilon b \ll 1$.
Higher loops are negligible if the relevant coupling $h$
 remains small at the mass scale set by $X$, $h \ll X$.
 This boils down to the inequality $h \ll \epsilon b \mu$.
 All in all there is a trustable non-supersymmetric vacuum if
\begin{equation}
\label{marg2}
\left(\frac{h}{\mu}\right)^2 
\ll 
\epsilon  \hat b \ll 
\frac{h}{\mu}
\end{equation}

\tocless\subsubsection{Marginal and relevant couplings}
One can consider more generic situations, where both $M^{(1)} \neq 0$ and $M^{(2)} \neq 0$.
This case has been discussed in \cite{Amariti:2009kb}, by considering the superpotential 
\begin{equation}
W= f X + h X \phi_1^2 + m \phi_1 \phi_2+
 \lambda X^2 \phi_3^2 + \mu \phi_3 \phi_4
\end{equation}
that combines (\ref{rel}) and the R-symmetric part of (\ref{marg}).
The analysis shows that there are regimes of masses 
and coupling such that this theory has an unstable 
origin, a long lived metastable R-symmetry and supersymmetry breaking minimum and 
a runaway in the large field region.
\\
\\
{\bf Summarizing}:
the general message that can be extracted from the analysis 
is that (meta)stable supersymmetry breaking in 3d 
WZ models is reliable at perturbative level if
R-symmetry is broken, either explicitly or spontaneously
\footnote{This analysis is reminiscent of the 4d results of  
\cite{Nelson:1993nf,Shih:2007av,Komargodski:2009jf}.
It should be interesting to further study the relation between the 3d and the 4d results for a full understanding of
the role of the R-symmetry in 3d $\mathcal{N}=2 \rightarrow \mathcal{N}=0$ supersymmetry breaking.}.
Spontaneous R-symmetry breaking is possible
whenever there are marginal couplings
involving the pseudomodulus in the superpotential.
On the other hand the explicit breaking is often necessary for models with relevant couplings. Actually there are situations 
with spontaneous R-symmetry breaking also for models with relevant couplings, depending on the 
R-charge assignations to the chiral superfields \cite{Amariti:2009kb}.

\subsection{Monopole quivers}

Here we introduce the notion
of monopole quivers. They have been originally defined in \cite{Amariti:2017gsm}, in the 
study of  the dimensional reductions of 4d dualities from the  D-brane perspective.

Consider a 4d gauge theory, here $U(N_c)$ SQCD
with $N_f=N_a$.
When this theory is dimensionally reduced on a circle of radius $r$
it gives raise to a 3d effective gauge theory with the same 
field content of the 4d parent, gauge coupling $g_3$ and with a further interaction,
due to the Kaluza-Klein (KK) monopole,
$W_{KK} = \eta e^{\Sigma_{N_c}- \Sigma_{1}}$
\cite{Aharony:2013dha}, where $\eta \propto e^{-\frac{1}{r g_3^2}}$.
At a generic point in the Coulomb branch the superpotential
is described by a further contribution due to the BPS monopoles
\begin{equation}
\label{WCB}
W =
W_{BPS} + W_{KK} = \sum_{i=1}^{N_c-1} 
e^{\Sigma_{1}- \Sigma_{i+1}}
+
\eta e^{\Sigma_{N_c}- \Sigma_{1}}
\end{equation} 
One can consider more complex situations, where the 
gauge groups is broken by the real scalars in the vector multiplet into a product of $r$ $U(n_i)$ factors, with 
$n_1+n_2+\dots+ n_r  =N_c$ and when real scalars in the background flavor symmetry set $f_i$ massless flavors 
at each $U(n_i)$ factor, such that 
$f_1+f_2+\dots+ f_r  =N_f$.
In this case an AHW superpotential, similar to (\ref{WCB})
is generated 
\begin{equation}
\label{Wmq}
W = \sum_{i=1}^{r} \lambda_i
e^{\Sigma_{n_i}^{(i)} - \Sigma_{1}^{(i+1)}}
\end{equation} 
with $n_{r+1} = n_1$. 
This construction is the prototypical
example of the monopole quiver studied in 
\cite{Amariti:2017gsm}. The definition is generalized to include 
real gauge groups, but we will not consider this possibility here.
The coupling constants $\lambda_i$ have a simple interpretation
in the brane setup. In this case the gauge groups live on D3 branes
separated along the compact direction by D1 branes, signaling the 
monopole interactions.
The couplings $\lambda_i$ in (\ref{Wmq}) are related to the distance 
between the D3 branes in this setup.

A further generalization, relevant for our 
analysis, includes 
an asymmetric content of fundamentals/anti-fundamentals and CS terms.
Moreover these monopole quivers can be defined also for pure
3d models, without requiring the presence of an extra circle.
In general they give raise to linear rather then circular 
monopole quivers with unitary gauge groups.

\begin{figure}
\begin{center}
\includegraphics[width=10cm]{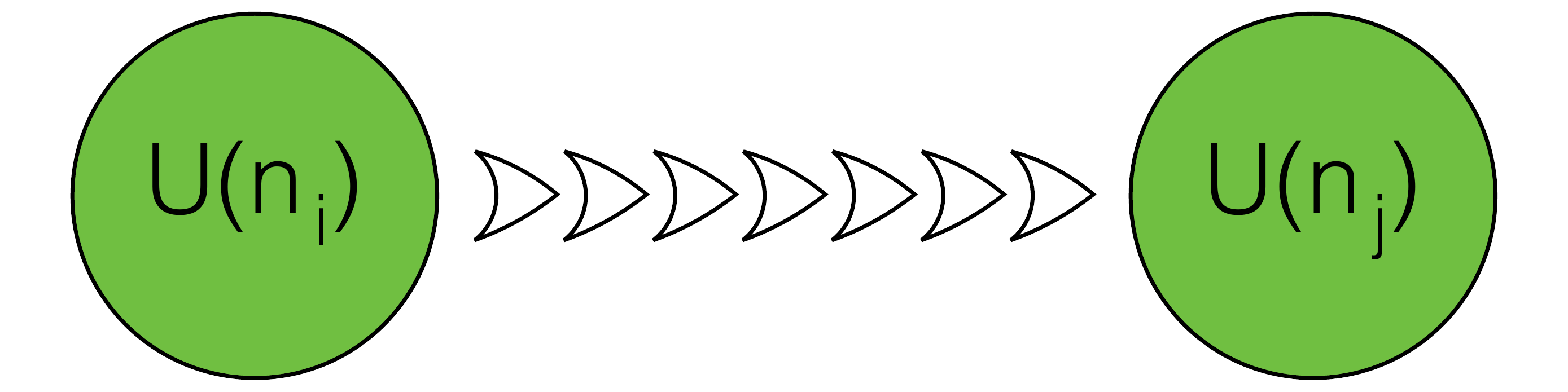}
\caption{Graphical representation of an AHW 
interaction between two 
gauge nodes. This represents one of the building blocks
of the monopole quivers.}
\label{ex}
\end{center}
\end{figure}

In order to provide a graphical representation of 
the monopole quiver we refer to the interaction 
$
e^{\Sigma_{n_i}^{(i)} - \Sigma_{1}^{(j)}}
\equiv  \widetilde  T_{n_i} T_{n_j}
$
as an oriented sequence of  triangles connecting the two gauge nodes
$U(N_i)$ and $U(N_{j})$. The orientation
of the triangle is fixed as in Figure \ref{ex}.

\subsection{A  survey of  3d $\mathcal{N}=2$ dualities}
\label{subsec:dualities}
Here we survey the 3d $\mathcal{N}=2$ dualities
relevant for our discussion. We restrict ourself to $U(N_c)_k$ gauge factors with $N_f$ fundamentals and $N_a$ anti-fundamentals.
We will be interested in two broad families of 3d $\mathcal{N}=2$ dualities: mirror symmetry and Seiberg like dualities.

\subsubsection*{Mirror symmetry}

Here we review mirror symmetry
for a $U(1)$ gauge theory with a pair of charged fields, $Q$ and $\tilde Q$ with
opposite charge, $+1$ and $-1$ respectively.
This model, usually referred to as 3d $\mathcal{N}=2$  SQED
can be equivalently described in terms of three chiral fields
$X$, $Y$ and $Z$ interacting through the superpotential 
$W = X Y Z$. The duality maps the field $X$ to the gauge invariant combination 
$Q \tilde Q$ in SQED, while the other two singlets $Y$ and $Z$ correspond to the monopole $T$ and 
the anti-monopole $\widetilde T$ in SQED.

By assigning a large real mass to one of the fields in the electric theory a new duality emerges, relating  
$U(1)_\frac{1}{2}$ with a field $Q$ to a singlet.
Such a singlet 
corresponds to the monopole in the electric phase.
This is the minimal version of mirror symmetry, originally discussed by \cite{Dorey:1999rb,Tong:2000ky}.

In the rest of the paper we will make a large use of these abelian mirror symmetries.
The nature of the interactions in the monopole quivers 
allows to use abelian mirror symmetry separately at the various nodes
\footnote{Observe that a similar procedure has been used in \cite{Aharony:2013dha}, by considering the mirror 
of a single abelian sector in a more complicated model.
This was possible because the different gauge sectors interacted only
through AHW superpotentials.}.
For example let us consider a pair of SQEDs connected by an 
AHW interaction
\footnote{From now on we suppress the couplings $\lambda_i$ in (\ref{Wmq})
by absorbing them in the definition of the monopoles, but we 
turn the coupling on in the WZ dual models, treating the fields as free.}
\begin{equation}
\label{AHW1}
W_{AHW} 
 =
 T_1 \widetilde T_2
\end{equation}
This is a simple example of monopole quiver with two gauge groups.
In this case we can apply mirror symmetry separately on each node and obtain the following duality map
\begin{equation}
Q_1 \widetilde Q_1 \leftrightarrow X_1,\quad
Q_2 \widetilde Q_2 \leftrightarrow X_2,\quad
T_1 \leftrightarrow Y_1,\quad
T_1 \leftrightarrow Y_2,\quad
T_2 \leftrightarrow Z_1,\quad
\widetilde T_2 \leftrightarrow Z_2
\end{equation}
where $Q_1$ ($Q_2$) and $\widetilde Q_1$ 
($\widetilde Q_2$) are the flavors of the first (second)
abelian gauge group.
The dual phase then corresponds to a pair of WZ 
models of the XYZ type discussed above, 
coupled by a massive interaction $Y_1 Z_2$.
The final superpotential in this case is
\begin{equation}
W = h_1 X_1 Y_1 Z_1 + h_2 X_2 Y_2 Z_2 + m \, Y_1 Z_2
\end{equation}
Observe that we could have modified the superpotential 
(\ref{AHW1}) as
\begin{equation}
\label{AHW2}
W_{AHW} 
 =
 T_1 \widetilde T_2 + T_2 \widetilde T_1
\end{equation}
This modification implies a circular shape for the monopole quiver, with the net 
effect to add a new reversed arrow to the one connecting the two gauge nodes in Figure \ref{ex}.
In this case the superpotential of the dual WZ model 
is
\begin{equation}
W = h X_1 Y_1 Z_1 + h X_2 Y_2 Z_2 + m (Y_1 Z_2+Y_1 Z_2)
\end{equation}

\subsubsection*{Seiberg dualities and its generalizations}

To complete this section we give a brief overview of
three dimensional non-abelian dualities generalizing 
4d Seiberg duality.
Three dimensional non-abelian dualities assume different forms, depending on the 
presence of CS terms in the action and of possible superpotentials involving monopole operators.
Here we list the possible dualities by specifying the superpotential, the matter 
content and the possible presence of CS terms in the electric theory.
Moreover we refer to the papers in which they have been first derived.
\begin{center}
\begin{tabular}{|c|c|c|c|c||}
\hline
$W_{ele}$& $k$    &  $N_a-N_f$ &   Reference\\
\hline
$W=0$        & $k=0$  &       $=0$     & \cite{Aharony:1997gp} \\
\hline
$W=0$        & $k \neq 0$  &       $=0$     &
\cite{Giveon:2008zn}\\
\hline
$W=0$        & $k\neq0$  &       $\neq 0$     &\cite{Benini:2011mf}\\
\hline
$W=T \widetilde T$        & $k=0$  &       $=0$     &
\cite{Aharony:2013dha}\\
\hline
$W=T + \widetilde T$        & $k=0$  &       $=0$     &
\cite{Benini:2017dud}\\
\hline
$W=T^2 + \widetilde T^2$        & $k=0$  &       $=0$     &
\cite{Benini:2017dud}\\
\hline
$W=T$        & $k \neq 0$  &       $\neq 0$     &
\cite{Benini:2017dud}\\
\hline
$W=T $        & $k=0$  &       $=0$     &
\cite{Benini:2017dud}\\
\hline
$W=T^2 $        & $k=0$  &       $=0$     &\cite{Amariti:2018gdc}\\
\hline
$W=T^2 $        & $k\neq 0$  &       $\neq 0$     &\cite{Amariti:2018gdc}\\
\hline
\end{tabular}
\end{center}
The monopole operators $T$ and $\widetilde T$ in the table
refer here to the ones with flux $(\pm 1,0,\dots,0)$.
The CS level can be integer or semi-integer depending on
the parity of $N_a-N_f$.
We do not provide further explanations of these dualities here 
and refer the reader to the original papers for definitions
and details.

\section{Supersymmetry breaking from monopole quivers }
\label{sec:exDSB}

This is the main section of the paper, where we 
study various monopole quivers leading to spontaneous supersymmetry breaking 
in the IR.
We organize the section by discussing examples of increasing complexity by enlarging 
the number of gauge groups and charged fields.
We restrict to monopole quivers with abelian gauge factors, with matter content given by
one flavor (with $k=0$) or one (anti)-fundamental (with $k=\frac{1}{2}$). 
Mirror symmetry can be performed at each node separately, because of the AHW interactions among the nodes
of the quiver.
This allows us to construct WZ models starting from monopole quivers.
Massive or more general superpotential deformations of the original quivers
can be added to the discussion without spoiling the duality.
By considering complex masses for the favors and linear or quadratic monopole deformations we  construct WZ models 
with the structure discussed in \cite{Amariti:2009kb} that lead to (meta)stable supersymmetry breaking  in the IR.
\subsection{Model I: Two abelian gauge groups}
This is the simplest example that we consider.
There are two $U(1)$ gauge groups.
The $U(1)$ factor associated to the first node has a CS 
term at level $k=\frac{1}{2}$ and there is just one charged field at charge $-1$.
The second $U(1)$ factor  has one field at charge $1$ and one at charge $-1$.
There is  an AHW superpotential, that connects the two gauge nodes as 
represented in Figure \ref{models}, of the form
\begin{equation}
W_{AHW} = T_1 \widetilde T_2 
\end{equation}
We deform the model by two superpotential terms, one consists 
of a quadratic term for the monopole, $W_{mono} = T_2^2$
while the other one is a mass term, $W_{mass} =  m Q \tilde Q$.
The final form of the superpotential is
\begin{equation}
W = W_{AHW}+ W_{mono} +W_{mass}  = 
T_1 \widetilde T_2 
+T_2^2
+
m \, Q \tilde Q
\end{equation}
This model can be dualized by applying mirror symmetry separately on both gauge nodes.
Here mirror symmetry maps the monopoles and the singlets of the electric theory 
to singlets in the dual WZ model.
This mapping is
\begin{equation}
\label{dmap}
T_1 \leftrightarrow W, \quad
T_2 \leftrightarrow Z, \quad
\tilde T_2 \leftrightarrow Y, \quad
Q \tilde Q \leftrightarrow X
\end{equation}
The mirror theory has superpotential
\begin{equation}
W = f X + h XYZ + m_1 Y^2 + m_2 ZW
\end{equation}
In the regime $m_1 \gg m_2 \gg f^\frac{2}{3}$ we 
integrate out the field $Y$ and arrive to the superpotential 
 \begin{equation}
W = f X + \hat h X^2 Z^2  + m_2 Z W
\end{equation}
where $h \propto \frac{h^2}{m}$ and $[\hat h ] =0$.
This superpotential corresponds to the one in formula 
(\ref{rel})
and it breaks supersymmetry if the parameters are chosen in the regime discussed in sub-section \ref{revSB}.
\subsection{Model II: Adding a singlet}
A second  model is obtained by adding
a massive singlet to the previous quiver and
by considering the superpotential
\begin{equation}
W = T_1 \widetilde T_2  +T_2 
+ \lambda  \, Q M \tilde Q + m\, M^2
\end{equation}
The mirror dual is a WZ model with superpotential
\begin{equation}
W = h \,X Y Z + f \,Y +  m_X\, X M + m\, M^2 + \mu \, Z W
\end{equation}
with the duality map still given by (\ref{dmap}).
In the regime $m_X,m \gg \mu$
we can integrate out the massive field $X$ and obtain
the superpotential
\begin{equation}
W = \hat h  \, Y^2 Z^2  +f \,Y + \mu \,Z W
\end{equation}
that corresponds to (\ref{rel})
and breaks supersymmetry if the  parameters
are chosen as discussed in sub-section \ref{revSB}.
\subsection{Model III: A circular monopole quiver}
\label{III}
Here we consider a monopole quiver with two $U(1)$ 
gauge groups, each one with a pair of charged fields,
with opposite charge $\pm 1$.
The associated quiver is given in Figure \ref{models}. 
The AHW interactions is
 \begin{equation}
  \label{circahw}
W_{AHW} = T_1 \tilde T_2 + T_2 \tilde T_1
\end{equation}
and we deform the model by the superpotential interactions
\begin{equation}
W_{def} = m Q \tilde Q + h (P \tilde P)^2
\end{equation}
We can apply mirror symmetry on both nodes
and obtain a WZ model.
After defining the mapping 
\begin{eqnarray}
\label{mapcirc}
 &&
 Q \tilde Q \leftrightarrow X_1, \quad
 T_1 \leftrightarrow Y_1, \quad
 \widetilde T_1\leftrightarrow Z_1
 , \quad
 P \tilde P \leftrightarrow X_2, \quad
 T_2 \leftrightarrow Y_1, \quad
  \widetilde T_2 \leftrightarrow Y_1
\end{eqnarray}
the superpotential of the dual WX model is
\begin{eqnarray}
W = f X_1 + h_1 X_1 Y_1 Z_1 + m( Y_1 Z_2 +Y_2 Z_1)
+ h_2 X_2 Y_2 Z_2 + \mu X_2^2
\end{eqnarray}

Following the general discussion of \cite{Giveon:2009bv,Amariti:2009kb} here a perturbative regime is 
possible if we add a term $h(Q \tilde Q)^2$ to the electric superpotential. This boils down 
to a term $\epsilon X^2$ in the dual phase, that breaks the R-symmetry and allows the existence of
a perturbative regime.

In appendix \ref{appB} we study the behavior of the non-supersymmetric state. 
We find that this model has a  perturbatively accessible 
 metastable non-supersymmetric vacuum, with a parametrically large lifetime, 
 if the parameters are chosen properly.
\subsection{Model IV: A circular quiver with an extra singlet}
\label{IV}

Here we add a further singlet $M$ to the model discussed above and consider the superpotential
\begin{equation}
W_{def} = m Q \tilde Q + h (Q \tilde Q)^2
+ M P \tilde P + m_M^2 M
\end{equation}
This superpotential has to be added to the AHW superpotential 
(\ref{circahw}).
Mirror symmetry on both nodes gives
\begin{equation}
W = f X_1 + \epsilon X_1^2 + 
X_1 Y_1 Z_1 + m( Y_1 Z_2 +Y_2 Z_1)
+ X_2 Y_2 Z_2 + \mu X_2 M + m_M^2 M
\end{equation}
where the duality map is again given by 
(\ref{mapcirc}).
We consider the regime $m_M,\mu  \gg m$ such that, after integrating out the massive fields $M$ and $X_2$,
we arrive at
\begin{equation}
W = f X_1 + \epsilon X_1^2 + 
X_1 Y_1 Z_1 + m( Y_1 Z_2 +Y_2 Z_1)
+ \frac{m_M^2}{\mu} Y_2 Z_2 
\end{equation}
This superpotential  corresponds to the one in formula 
(\ref{marg})
if the parameters are chosen such that 
$\epsilon = \frac{m_M^2}{\mu}$,
and it breaks supersymmetry 
in the regime (\ref{marg2}).
\subsection{Model V: Three abelian factors}
Here we discuss a monopole quiver with two gauge groups
$U(1)_{1/2} \times U(1)_0 \times  U(1)_{1/2}  $.
The $U(1)_{1/2} $  nodes  have $N_f^{(1)}=1$ and $N_a^{(1)}=0$ each
while the node $U(1)_0$  has $N_f^{(2)}=1$ and $N_a^{(2)}=1$. 
The quiver is shown in Figure \ref{models}.
The AHW interactions is
 \begin{equation}
W_{AHW} = T_1 \tilde T_2 +T_2 \widetilde T_3
\end{equation}
and we deform the model by the superpotential interactions
\begin{equation}
W_{def} = m Q \tilde Q 
\end{equation}
We use mirror symmetry on each node obtaining a WZ model with superpotential
\begin{equation}
W = m_T(T_1 \widetilde T_2 + T_2 \widetilde T_3) 
+ m X + h X T_2 \tilde T_2
\end{equation}
We could also have added a monopole deformation $T_1^2$
that would have led to the superpotential
\begin{equation}
W =  m_T T_2 \widetilde T_3 + m X + \lambda X^2 T_2^2
\end{equation}
corresponding to the one in formula 
(\ref{rel}).
\subsection{Model VI: Four abelian factors}
\label{modVI}
\begin{figure}[htpb]
\begin{tabular}{|c|c|}
\hline
\includegraphics[width=7cm]{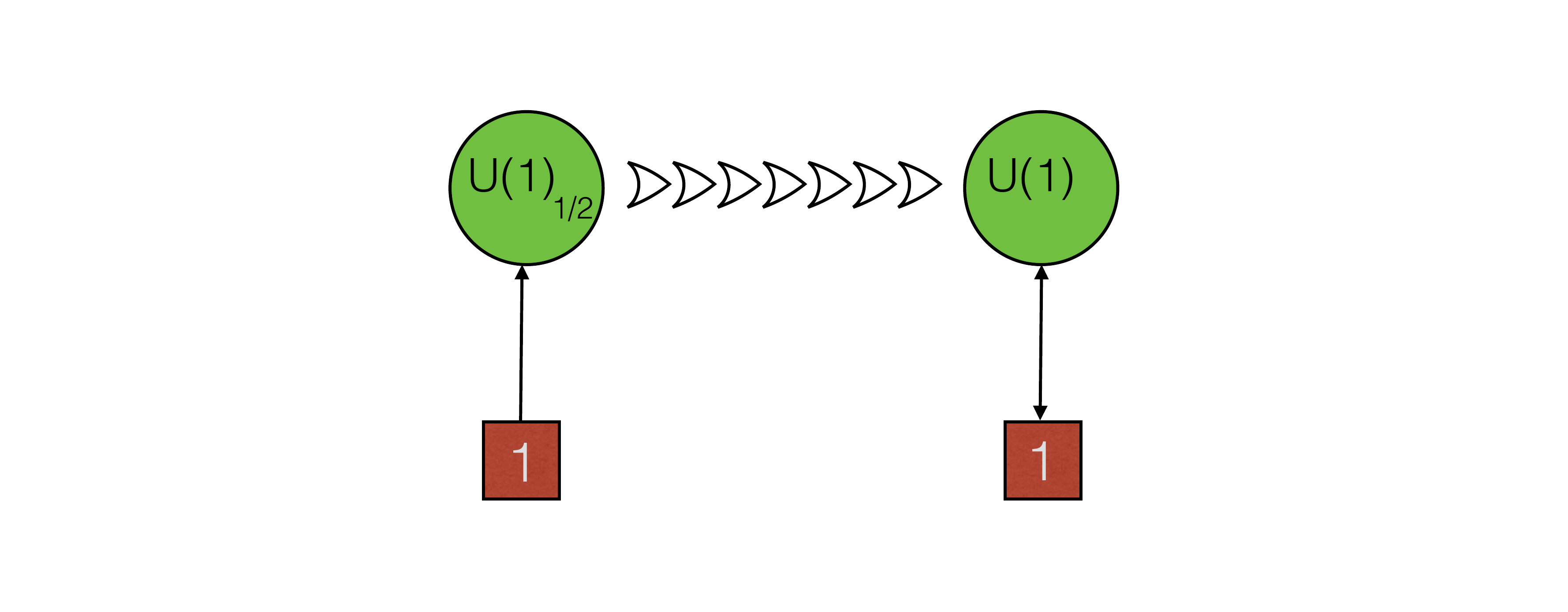}
&
\includegraphics[width=7cm]{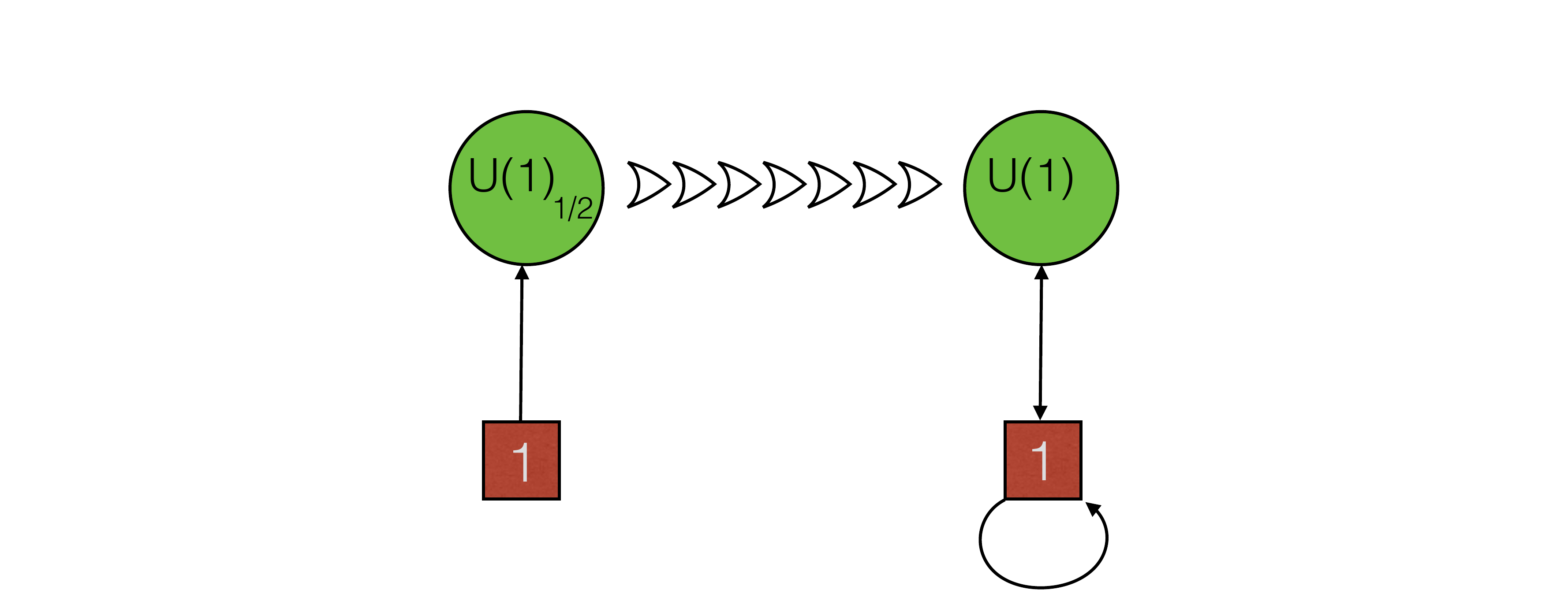}
\\
\hline 
Model I
&
Model II \\
\hline
\includegraphics[width=7cm]{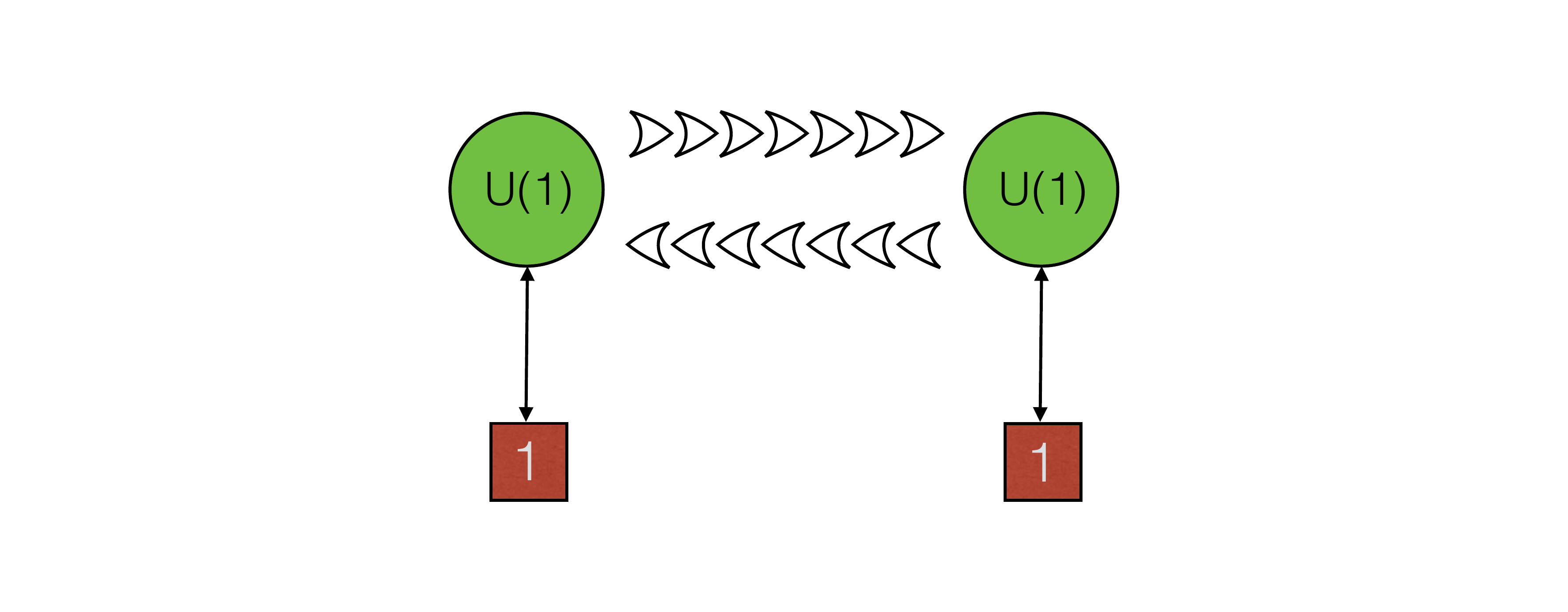}
&
\includegraphics[width=7cm]{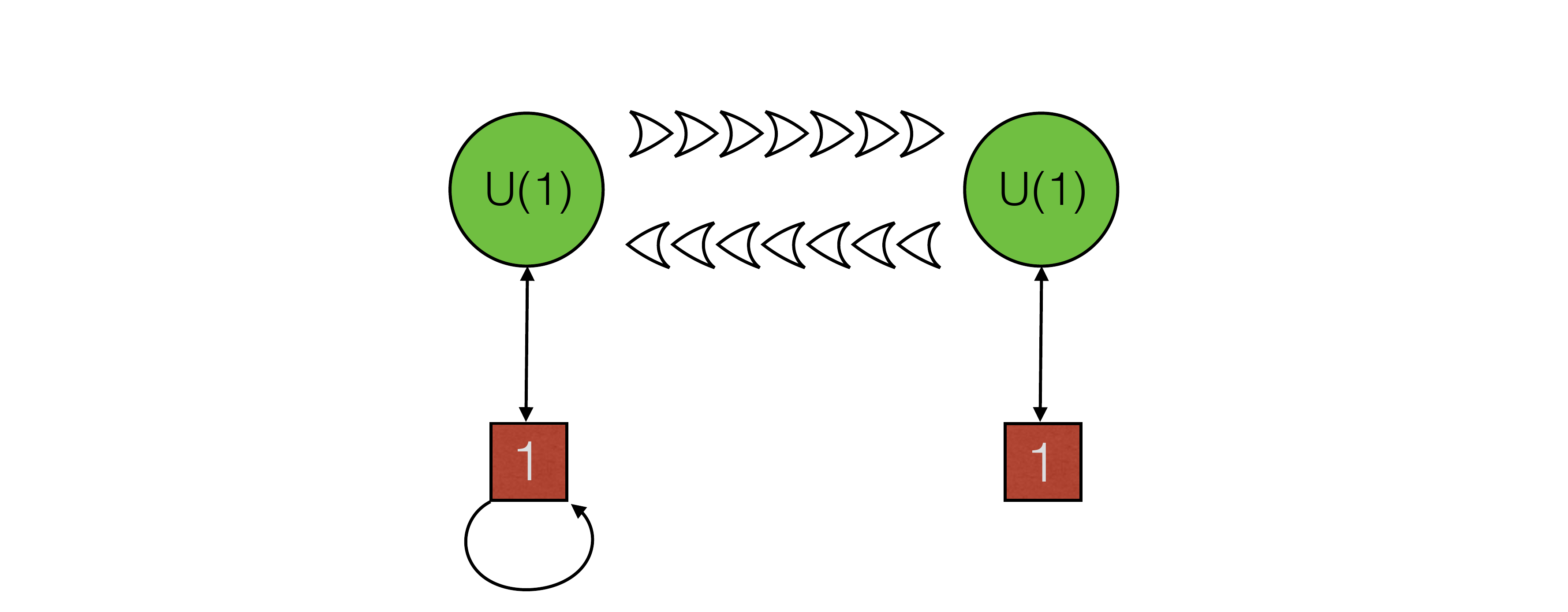}
\\
\hline 
Model III
&
Model IV \\
\hline
\includegraphics[width=7cm]{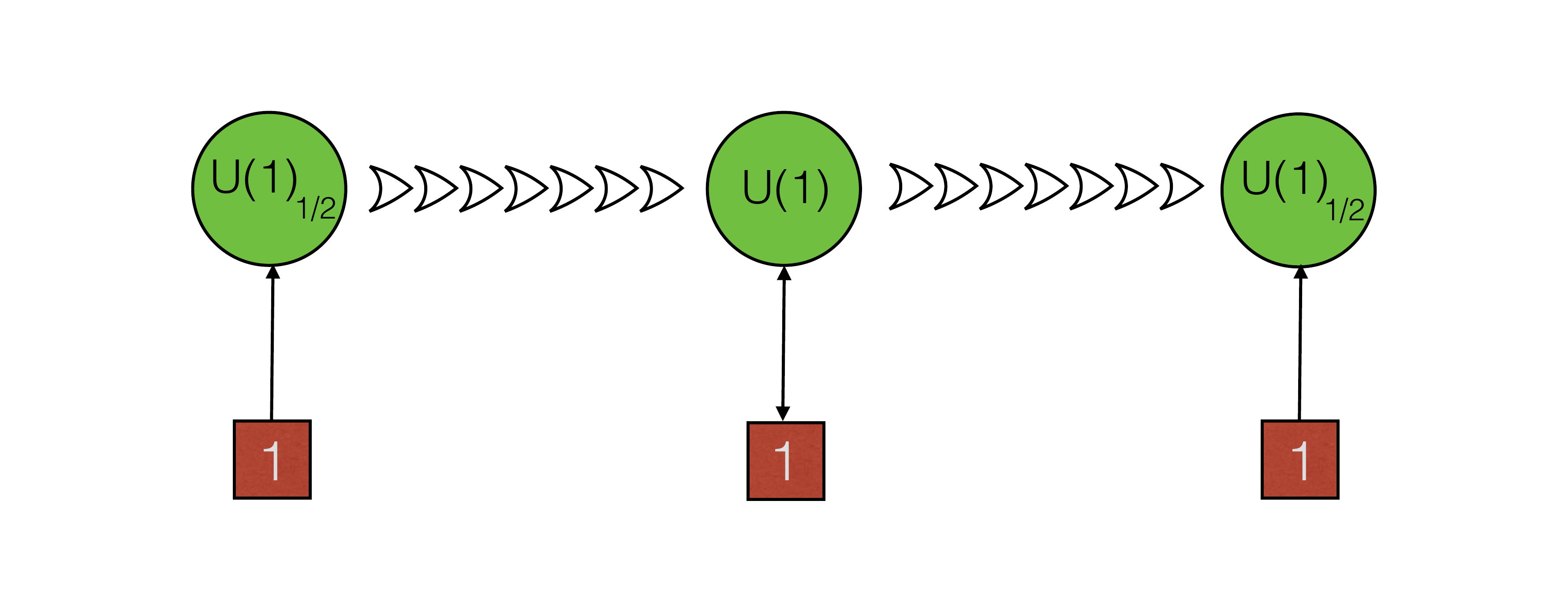}
&
\includegraphics[width=7cm]{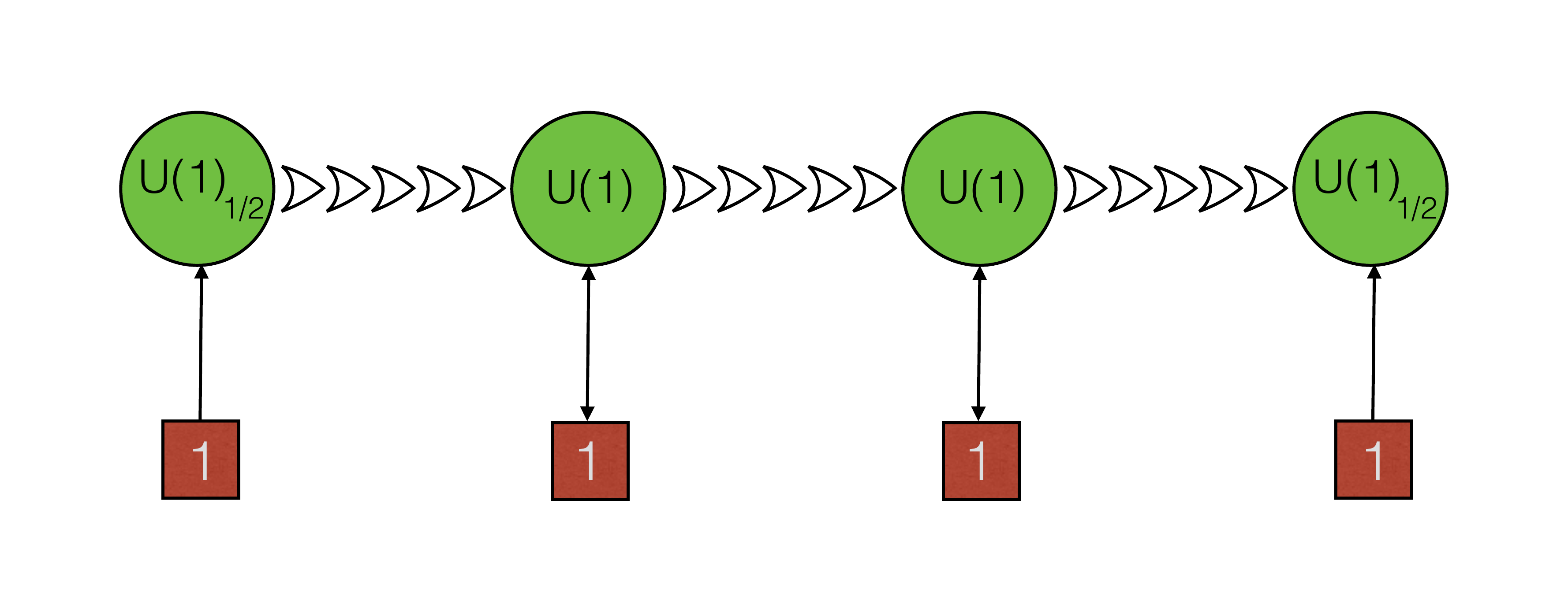} 
\\
\hline 
Model V
&
Model VI \\
\hline
\end{tabular}
\caption{Summary of the monopole quivers studied in the paper.}
\label{models}
\end{figure}

Here we consider a monopole quiver with gauge group
$U(1)_\frac{1}{2} \times U(1) \times U(1) \times U(1)_\frac{1}{2}$
as in Figure \ref{models}.
The AHW superpotential is 
 \begin{equation}
W_{AHW} = T^{(1)} T_+^{(2)} +  T_-^{(2)} T_+^{(3)} +T_-^{(3)} T^{(4)}
\end{equation}
and we add the deformation
 \begin{equation}
W_{def}  =  \mu Q \tilde Q + h (P \tilde P)^2 
\end{equation}
Applying  mirror symmetry on each node we arrive at a WZ model
with superpotential
 \begin{equation}
W= m (\phi_1 \phi_3+\phi_2 \phi_4+\phi_5 \phi_6) + h_X X \phi_1 \phi_2 + f X + h_Y Y \phi_4 \phi_5 + m_Y Y^2
\end{equation}
In this case a perturbative regime for the tree level
non-supersymmetric state can be realized by adding a term $h(Q \tilde Q)^2$ to the electric superpotential. This boils down 
to a term $\epsilon X^2$ in the dual phase, that breaks the R-symmetry and allows the existence of
a perturbative regime.

In appendix \ref{appC} we show that this model has a  perturbatively accessible 
 metastable non-supersymmetric vacuum, with a parametrically large lifetime, 
 in the opportune regime of parameters.

\section{Conclusions}
\label{ref:conclusions}
In this paper we have supported the claim that supersymmetry breaking in 3d $\mathcal{N}=2$ gauge theories is 
quite generic in large classes of monopole quivers.
We restricted our attention to quivers with only abelian
nodes because in such cases the analysis is simplified by the use of 
 mirror symmetry.
The presence of  AHW monopole interactions between the various gauge groups justified the use of
abelian mirror symmetry separately at each node.
In this way the low energy descriptions
consist of WZ models, such that, for opportune choices of couplings,
 the SUSY broken phase becomes perturbatively accessible.
 
 In the analysis we did not find any dynamical mechanism fixing the
 hierarchy of couplings that  leads to the existence of 
 stable and calculable non-supersymmetric states. 
This is a crucial issue that requires the study of 
the UV completions of the models treated here.
We hope to come back to this problem in the future.

It should be also interesting to generalize our analysis to 
non-abelian gauge groups.
In some cases supersymmetry breaking is expected 
because the Seiberg-like dual phases, in presence of 
monopole superpotential deformations, are
of the type discussed in sub-section \ref{subsec:dualities}, and there is 
a negative rank for the dual gauge group.
This is a signal
of a supersymmetry breaking state in the spectrum.
It is not possible to use abelian mirror symmetry
 in such cases, but 
it may be interesting to study these models along 
the analysis of \cite{Giacomelli:2017vgk}.

Another interesting case that may deserve a deeper analysis
corresponds to the circle compactification of SQCD.
In presence of a light mass deformations in the 4d IR free window the ISS \cite{Intriligator:2006dd} 
mechanism is at work. It should be interesting to study its generalization in 3d in presence of 
a KK monopole superpotential.
Furthermore it should be possible to find new SUSY breaking models by increasing 
the number of gauge nodes, both for cases of circular quivers and 
for cases with linear ones. We leave such an analysis to the
future.

\section*{Acknowledgments}
This work is supported in part by Italian Ministero dell'Istruzione, Universit\`a 
e Ricerca (MIUR) and Istituto Nazionale di Fisica Nucleare (INFN) through 
the "Gauge Theories, Strings, Supergravity" (GSS) research project. 
We thank the Galileo Galilei Institute for Theoretical Physics for 
the hospitality and the INFN for partial support during the early stages of this work.

\appendix
\section{Tools in 3d WZ}
\label{appendix}
\subsection{One loop effective potential}

In the body of the paper we discussed many 
gauge theories dual to WZ models and studied 
supersymmetry breaking in the latters.
At tree level supersymmetry is broken because the
F-terms cannot be solved simultaneously.
The configuration with minimal energy breaks supersymmetry 
at tree level.
Often this is not just a vacuum state, but 
there are scalar flat directions.
Such moduli are
actually pseudomoduli and they can 
acquire loop corrections.
These corrections can either lift the moduli, isolating a single vacuum 
state or provide a negative squared mass, destabilizing the vacuum.
The one loop effective potential  is
given by the CW formula
\begin{equation}
\label{CW}
V_{CW} = 
-\frac{1}{12 \pi} 
\text{STr}  |\mathcal{M}|^3
\equiv
-\frac{1}{12 \pi} 
\text{Tr} \left(|\mathcal{M}_B|^2 -  |\mathcal{M}_F|^2\right)
\end{equation}
where $\mathcal{M}_B$ and 
$\mathcal{M}_F$ are the tree level bosonic 
and fermionic mass matrices respectively.

\subsection{Bounce action}

The lifetime of a metastable non-supersymmetric state in 3d can be estimated through the 
calculation of the bounce action $S_B$  for a triangular barrier. This is a sensible estimation
if the two vacua are very far in field space (or in presence of  a runaway potential).
The result has been obtained in \cite{Amariti:2009kb}, along the lines of 
\cite{Duncan:1992ai}, computing $S_B$ as the difference between the tunneling configuration
and the metastable vacuum in the euclidean action.
The analysis has been performed in terms of a single field $\phi$ decaying from the false vacuum $\phi_F$ to the 
true vacuum $\phi_T$.
The result is expressed in terms of a 
dimensionless parameter $c \equiv -\lambda_T/\lambda_F$, where
\begin{equation}
\lambda_F
=
\frac{V_{max} - V_{F}}{\phi_{max} - \phi_F}
\equiv
\frac{\Delta V_F}{\Delta \phi_F},
\quad
\lambda_T
=
\frac{V_{max} - V_{T}}{\phi_{max} - \phi_T}
\equiv
-\frac{\Delta V_T}{\Delta \phi_T}
\end{equation}
and $F$ and $T$ stay for true and false.
By using these definitions the bounce action is
\begin{equation}
S_B = \frac{16 \sqrt 6\pi}{5}
\frac{1+c}{(3+2c-3(1+c)^{2/3})^{3/2}}
\sqrt \frac{\Delta \phi_F^6}{\Delta V_F}
\end{equation}

\section{Susy breaking for Model III}
\label{appB}
In this section we study the stability of non-supersymmetric vacuum 
claimed in section \ref{III}.
For simplicity we re-organize the superpotential as
\begin{equation}
W =  h  X \phi_1 \phi_2 -f X+ h Y \phi_3 \phi_4 + h \mu (\phi_1 \phi_3 + \phi_2 \phi_4) + \frac{1}{2} h \mu (\epsilon_X X^2 + \epsilon_Y Y^2)
\end{equation}
where the parameters $h$ and $\mu$ have mass dimensions $\frac{1}{2}$,
the supersymmetry breaking scale has mass dimensions $\frac{3}{2}$ and the parameters
$\epsilon_X$ and $\epsilon_Y$ are dimensionless.

At small $X$ the theory has a non-supersymmetric state in the regime $\epsilon_X \ll 1$.
If we compute the bosonic masses for the fields $Y$ and $\phi_i$ in this region we have
\begin{eqnarray}
m_Y^2 = h^2 m^2 \epsilon_Y^2,\quad
m_{\phi_{i}}^2 = \left \{ \frac{h}{2} (a_\pm + \sqrt {a_\pm^2-b_\pm})
,\quad
\frac{h}{2} (a_\pm - \sqrt {a_\pm^2-b_\pm})
\right \}
\end{eqnarray}
with
\begin{eqnarray}
\label{abpm}
a_{\pm} = 
2 h \mu ^2+h X^2\pm (f-h \mu  X \epsilon_X), \quad
 b_{\pm} = 
  4 h \mu ^2 ( h \mu ^2\pm (f-h \mu  X \epsilon_X))
 \end{eqnarray}
The tree level scalar potential for the field $X$ together to the one loop 
correction from the CW potential is 
\begin{equation}
 V(X)
\simeq
 | 8 \pi \mu^2 \epsilon_h |^2 |\epsilon_X  X  -\mu \, \epsilon _f |^2
-|8 \pi ^2 \mu ^4 \epsilon _f \epsilon _h^3 (\epsilon _f (X^2 (3 \epsilon_X^2-2)+8 \mu ^2)-16 \mu  X \epsilon _X)|
\end{equation}
where we expressed it in terms of the  dimensionless parameters $\epsilon_X\ll1$, $\epsilon_f \equiv \frac{f}{h \mu^2} \ll 1$ and  $\epsilon_h \equiv \frac{h}{8 \pi \mu}$.
The non-supersymmetric vacuum is placed at
\begin{equation}
  \langle X \rangle =    
\left | 
\frac{ \, \mu  \, (\epsilon _h-1)}{\frac{\epsilon _f}{\epsilon_X} \frac{\epsilon _h}{4} + \frac{\epsilon _X}{\epsilon_f} }
\right |
  \end{equation}
 In order to study the supersymmetry breaking scenario in such a vacuum 
 we need to impose the following constraints on the parameters
 \begin{itemize}
 \item The vacuum has to be close to the origin. This is necessary for having a parametrically large lifetime.
 This corresponds to the requirement
 \begin{equation}
|\langle X \rangle |\ll |\mu|
 \end{equation}
  On the other hand the supersymmetric vacuum,
  placed at 
$
|\langle X \rangle | =  \frac{|\epsilon_f|}{|\epsilon_X|} |\mu|
$,
 must be far in the field space
 \item The perturbative approximation has to be valid, i.e. higher loops must be suppressed.
  This corresponds to the requirement
 \begin{equation}
| h| \ll |\langle X \rangle |
 \end{equation}
 Observe that this fixes $\epsilon_h \ll 1$ as well
 \footnote{We further assume $\epsilon_h>0$, \emph{a posteriori} one can
 see that the opposite regime is not compatible with the various constraints.}.
 \item There should be no tachyons at the non-supersymmetric vacuum.
   This corresponds to the requirement
 \begin{equation}
|\langle X \rangle | \ll  \left | \frac {\mu(\epsilon_f \pm 1)}{\epsilon_X} \right | 
\end{equation}
 \end{itemize} 
 The various requirements are satisfied in the regime 
 \begin{equation}
 \label{regime}
2\pi \epsilon_h^2 \ll \frac{\epsilon_X}{\epsilon_f} \ll \frac{1}{4} \epsilon_h \ll 1
 \end{equation}
The squared mass at the vacuum is positive,
$
m_X^2 = 32 \pi ^2 \mu ^4  \epsilon _h^3 \epsilon _f^2 
$, and 
the bounce action is
\begin{equation}
S_B \propto  \frac{\epsilon_f^2}{16 \, \pi \, \epsilon_X^3 \epsilon_h} \gg 
\frac{1}{16\,  \pi \,  \epsilon_X  \epsilon_h} \gg 1
\end{equation}
On can then conclude that in this regime of parameters there is a
non-supersymmetric metastable
vacuum with a parametrically large lifetime.

\section{Susy breaking for Model VI}
\label{appC}

The other WZ model that has not been studied  in the literature yet
and that we have obtained in the body of the paper is model VI 
in sub-section \ref{modVI}.
Here we re-organize the superpotential as
\begin{equation}
W=h X \phi _1 \phi _2-f X+h \mu  \left(\phi _1 \phi _3+\phi _2 \phi _4+\psi_5 \psi _6\right)
+h Y \phi _4 \psi_5 +
\frac{1}{2} h \mu  (X^2 \epsilon _X+ Y^2 \epsilon _Y)
\end{equation}
where again the parameters $h$ and $\mu$ have mass dimensions $\frac{1}{2}$,
the scale $f$ has mass dimension  $\frac{3}{2}$ and the parameters
$\epsilon_X$ and $\epsilon_Y$ are dimensionless.

The non-supersymmetric state is still at small $X$ if $\epsilon_X \ll 1$.
The bosonic masses for the fields $Y$ and $\psi$ are supersymmetric and the
non-supersymmetric contribution to the CW potential for the field $X$ is
due to the masses of the fields $\phi_i$.
The masses for the bosonic components can be summarized as
\begin{eqnarray}
\label{mb}
m_{\psi_i}^2 = h^2 \mu^2, \,
m_Y^2 = h^2 \mu^2 \epsilon_Y^2,\, 
m_{\phi_i} =\frac{h}{2}  \left\{
(a_\pm + \sqrt {a_\pm^2-b_\pm})
,  \,
(a_\pm - \sqrt {a_\pm^2-b_\pm})
\right\}
\end{eqnarray}
with $a_{\pm}$ and $b_\pm$ given by
(\ref{abpm}).
This coincidence  and the equivalence of the 
tree level contribution to the scalar potential with the one of the model studied in 
appendix \ref{appB} are enough to claim that the rest of the analysis 
coincide with the one done there.
We can again conclude that in the regime 
(\ref{regime}) supersymmetry is broken in 
a perturbatively accessible metastable
vacuum with a parametrically large lifetime.

\end{document}